\documentclass[11pt]{article}%
\usepackage{amsmath}
\usepackage{amssymb}
\usepackage{amsfonts}

\usepackage{cite}
\usepackage{graphicx}
\usepackage{float}
\usepackage{longtable}
\usepackage[utf8]{inputenc}
\usepackage{hyperref}

\usepackage{algorithmicx}
\usepackage[ruled,vlined]{algorithm2e}
\SetAlgoCaptionSeparator{ }

\usepackage{fullpage}

\newcommand{\fref}[1]{Fig.~\ref{#1}}

\newcommand{\sref}[1]{Section~\ref{#1}}

\newenvironment{meth}[1][!h]
  {
   \begin{algorithm}[#1]%
  }{\end{algorithm}}

\newenvironment{algo}[1][!h]
  {
   \begin{algorithm}[#1]%
  }{\end{algorithm}}

\providecommand{\U}[1]{\protect\rule{.1in}{.1in}}
\begin{document}

\title{Opportunistic Entanglement Distribution for the Quantum Internet}
\author{Laszlo Gyongyosi\thanks{School of Electronics and Computer Science, University of Southampton, Southampton SO17 1BJ, U.K., and Department of Networked Systems and Services, Budapest University of Technology and Economics, 1117 Budapest, Hungary, and MTA-BME Information Systems Research Group, Hungarian Academy of Sciences, 1051 Budapest, Hungary.}
\and Sandor Imre\thanks{Department of Networked Systems and Services, Budapest University of Technology and Economics, 1117 Budapest, Hungary.}}
\date{}

\maketitle
\begin{abstract}
Quantum entanglement is a building block of the entangled quantum networks of the quantum Internet. A fundamental problem of the quantum Internet is entanglement distribution. Since quantum entanglement will be fundamental to any future quantum networking scenarios, the distribution mechanism of quantum entanglement is a critical and emerging issue in quantum networks. Here we define the method of opportunistic entanglement distribution for the quantum Internet. The opportunistic model defines distribution sets that are aimed to select those quantum nodes for which the cost function picks up a local minimum. The cost function utilizes the error patterns of the local quantum memories and the predictability of the evolution of the entanglement fidelities. Our method provides efficient entanglement distributing with respect to the actual statuses of the local quantum memories of the node pairs. The model provides an easily-applicable, moderate-complexity solution for high-fidelity entanglement distribution in experimental quantum Internet scenarios.
\end{abstract}

\section{Introduction}
\label{sec1}
Quantum entanglement has a central role in the quantum Internet \cite{ref1,ref2,ref3,ref4,ref5,ref6,ref7,ref8,ref9,ref10}, quantum networking \cite{ref11,ref12,ref13,ref14,ref18,ref19,ref20,ref21,ref22,ref23,ref24,ref25,ref26,ref27,ref28,ref29}, and long-distance quantum key distribution \cite{ref1,ref22,ref30,ref38}. Entanglement distribution is a crucial phase for the construction of the entangled core network structure of the quantum Internet. In a quantum Internet scenario, quantum entanglement is a preliminary condition of quantum networking protocols \cite{ref30,ref31,ref32,ref33,ref34,ref35,ref36,ref37,ref39,ref40,ref41,ref42,ref43,ref44,ref45,ref46,ref47}. Distant quantum nodes that share no quantum entanglement must communicate with their direct neighbors to distribute entanglement. To aim of entanglement distribution is to generate entanglement between a distant source node and a target node through a chain of intermediate quantum repeater nodes \cite{ref50,ref51,ref52,ref53,ref54,ref55,ref56,ref57,ref58,ref59,ref60,ref61,ref62}. The intermediate quantum repeater nodes receive the entangled states, store them in their local quantum memories \cite{ref38,ref48,ref49}, and apply a unitary operation (called entanglement swapping \cite{ref1,ref2,ref3,ref22}) to extend the range of quantum entanglement. Storing quantum entanglement in the quantum nodes' local quantum memories adds noise to the distribution process, since quantum memories are non-perfect devices \cite{ref36,a1} and the error probabilities evolve in time \cite{ref1, ref54,ref55,ref56,ref57,ref58,ref59}. As the error pattern of the evolution the quantum memories is predictable, the nodes can be characterized by a given storage success probability after a given time from the start of the storage. 

The fidelity of entanglement \cite{a2,a3,a4} is another critical parameter for entanglement distribution. In a quantum network with a chain of repeater nodes between a source and target nodes, for all pairs of entangled nodes (e.g., for nodes that share a common entanglement) a given lower bound in the fidelity of entanglement must be satisfied, otherwise the entanglement distribution fails \cite{ref1,a5}. The stored entangled states have a given amount of fidelity that is determined by the transmission procedure, such as the noise of the quantum channel, etc. The evolution of a given entangled system's fidelity parameter is time-varying in quantum memory, since it evolves through time, from the beginning of storage to the actual current time. Therefore, it is necessary to consider the predictability of the evolution of both the error patterns of the nodes' local quantum memories, and the evolution of the stored quantum states' fidelity of entanglement. In our model, using these parameters, we define an appropriate cost function for the realization of entanglement distribution. 

Here we define the method of opportunistic entanglement distribution for the quantum Internet. The proposed scheme utilizes a cost function that accounts for the error patterns of local quantum memories and also the evolution of entanglement fidelities. The opportunistic model defines distribution sets in the entangled quantum network of the quantum Internet. A distribution set contains those quantum nodes for which our cost function picks up a local minimum in comparison to the cost of the other nodes in the given distributing set. The distribution set selects a lowest-cost node from a given set of nodes to provide a maximal usability of stored entanglement. The cost function ensures that the nodes selected for entanglement distribution allow the lowest deviation in the entanglement fidelity from the start of storage, and that the behavior of the quantum memory error follows a predicted error model with respect to a given node pair. We also derive the computational complexity of the proposed method. The solution provides an easily-applicable, low-complexity solution for high-fidelity entanglement distribution in the quantum Internet.

The novel contributions of our manuscript are as follows:

\begin{enumerate}
\item \textit{We define the method of opportunistic entanglement distribution for the quantum Internet.}

\item \textit{The proposed opportunistic model defines distribution sets that are aimed to select those quantum nodes for which the cost function picks up a local minimum. The cost function utilizes the error patterns of the local quantum memories and the predictability of the evolution of the entanglement fidelities.} 

\item \textit{Our method provides efficient entanglement distributing with respect to the actual statuses of the local quantum memories of the node pairs.} 

\item \textit{We derive the computational complexity of the model.}
\end{enumerate}

This paper is organized as follows. In \sref{sec2} the preliminaries are summarized. \sref{sec3} defines the method, while \sref{sec4} proposes the results. Finally, \sref{sec5} concludes the results.

\section{Preliminaries}
\label{sec2}
\subsection{System Model}
The quantum Internet setting is modeled as follows \cite{ref8}. Let $V$ refer to the nodes of an entangled quantum network $N$, with a transmitter quantum node $A\in V$, a receiver quantum node $B\in V$, and quantum repeater nodes $R_i\in V$, $i=1,\dots ,q$. Let $E=\left\{E_j\right\}$, $j=1,\dots ,m$, refer to a set of edges between the nodes of $V$, where each $E_j$ identifies an $\text{L}_l$-level entangled connection, $l=1,\dots ,r$, between quantum nodes $x_j$ and $y_j$ of edge $E_j$, respectively. The entanglement levels of the entangled connections in the entangled quantum network structure are defined as follows.

\subsubsection{Entanglement Levels}
In a quantum Internet setting, an $N=\left(V,E\right)$ entangled quantum network consists of single-hop and multi-hop entangled connections, such that the single-hop entangled nodes\footnote{The $l$-level entangled nodes ${x,y}$ refer to quantum nodes $x$ and $y$ connected by an entangled connection ${\text{L}_l}$.} are directly connected through an $\text{L}_1$-level entanglement, while the multi-hop entangled nodes communicate through $\text{L}_l$-level entanglement. Focusing on the doubling architecture \cite{ref1,ref2,ref3} in the entanglement distribution procedure, the number of spanned nodes is doubled in each level of entanglement swapping (entanglement swapping is applied in an intermediate node to create a longer distance entanglement \cite{ref1}). Therefore, the $d{\left(x,y\right)}_{\text{L}_l}$ hop distance in $N$ for the $\text{L}_l$-level entangled connection between $x,y\in V$ is denoted by \cite{ref8}
\begin{equation} \label{eq1} 
d{\left(x,y\right)}_{\text{L}_l}=2^{l-1}, 
\end{equation} 
 with $d{\left(x,y\right)}_{\text{L}_l}-1$ intermediate quantum nodes between $x$ and $y$. Therefore, $l=1$ refers to a direct entangled connection between two quantum nodes $x$ and $y$ without intermediate quantum repeaters, while $l>1$ identifies a multilevel entanglement.

An entangled quantum network $N$ is illustrated in \fref{fig1}. The quantum network integrates single-hop entangled nodes (depicted by gray nodes) and multi-hop entangled nodes (depicted by blue and orange nodes) connected by edges. The single-hop entangled nodes are directly connected through an $\text{L}_1$-level entangled connection, while the multi-hop entangled nodes are connected by $\text{L}_2$ and $\text{L}_3$-level entangled connection.

The fidelity of entanglement of an $\text{L}_l$-level entangled connection $E\left(x,y\right)$ between $x,y\in V$ depends on the physical attributes of the quantum network.

\subsection{Terms and Definitions}
\subsubsection{Entanglement Fidelity}
Let
\begin{equation}
{\left| \beta _{00}  \right\rangle} ={\textstyle\frac{1}{\sqrt{2} }} \left({\left| 00 \right\rangle} +{\left| 11 \right\rangle} \right)
\end{equation}
be the target Bell state subject to be created at the end of the entanglement distribution procedure between a particular source node $A$ and receiver node $B$. The entanglement fidelity $F$ at an actually created noisy quantum system $\sigma $ between $A$ and $B$ is
\begin{equation}
F\left( \sigma  \right)=\langle  {{\beta }_{00}} |  \sigma |{{\beta }_{00}} \rangle ,
\end{equation}
where $F$ is a value between $0$ and $1$, $F=1$ for a perfect Bell state and $F<1$ for an imperfect state. Without loss of generality, in an experimental quantum Internet setting, an aim is to reach $F\ge 0.98$ over long distances \cite{ref1,ref3}.

Some properties of $F$ are as follows \cite{ref4,ref22}. The fidelity for two pure quantum states $| \varphi \rangle $, $| \psi \rangle $ is defined as 
\begin{equation} \label{eq1} 
F(| \varphi \rangle ,| \psi \rangle )=|\langle \varphi |\psi \rangle |^{2} . 
\end{equation} 
 The fidelity of quantum states can describe the relation of a pure state $|\psi \rangle $ and mixed quantum system $\sigma =\sum _{i=0}^{n-1} p_{i} \rho _{i} =\sum _{i=0}^{n-1} p_{i} | \psi _{i} \rangle  \langle \psi _{i} |$, as 
\begin{equation} \label{eq2} 
F(|\psi \rangle ,\sigma )=\langle \psi |\sigma |\psi \rangle =\sum^{n-1}_{i=0}{}p_i{|\langle \psi |{\psi }_i\rangle |}^2. 
\end{equation} 
 Fidelity can also be defined for mixed states $\sigma $ and $\rho $, as 
\begin{equation} \label{eq3} 
F(\rho,\sigma)={(\text{Tr}(\sqrt{\sqrt{\sigma }\rho \sqrt{\sigma }}))}^2=\sum_i{}p_i{(\text{Tr}(\sqrt{\sqrt{{\sigma }_i}{\rho }_i\sqrt{{\sigma }_i}}))}^2. 
\end{equation}

\section{Method}
\label{sec3}
Before giving the details of the algorithm, we briefly summarize the method of opportunistic entanglement distribution in a quantum Internet setting in Method 1. 

\setcounter{algocf}{0}
\begin{meth}
  \DontPrintSemicolon
\caption{Opportunistic entanglement distribution in the quantum Internet}

\textbf{Step 1}. Select a cheapest quantum node $i$. Generate entanglement between $i$ and the direct contacts of $i$. The entangled contacts of $i$ define the distributing set ${\rm {\mathcal D}}_{i} $ of $i$.

\textbf{Step 2}. From ${\rm {\mathcal D}}_{i} $, select a cheapest quantum node $j$. Generate entanglement between $j$ and the direct contacts of $j$ to define ${\rm {\mathcal D}}_{j} $. Select the cheapest node from ${\rm {\mathcal D}}_{j} $.

\textbf{Step 3}. Repeat steps 1-2 until the source and target nodes share entanglement.
 
\end{meth}

\subsection{Discussion}

In Step 1, distributing entanglement between $i$ and the direct contacts of $i$ defines the distributing set ${\rm {\mathcal D}}_{i} $, which can contain different levels of entangled contacts. In our opportunistic approach, the heterogonous entangled contacts leads to diverse hop-distances, specifically for an ${\rm L}_{l} $-level entanglement $l=1,\ldots ,r$, the $d\left(i,y\right)_{{\rm L}_{l} } $, the hop-distance between a source node $i$ and target node ${\rm {\mathcal D}}_{i} \in y$ from the distributing set ${\rm {\mathcal D}}_{i} $ is determined via \eqref{eq1} as $d\left(i,y\right)_{{\rm L}_{l} } =2^{l-1} $ \cite{ref1,ref2,ref3,ref8}. 

In Step 2, the next node $j$ from a set ${\rm {\mathcal D}}_{i} $ is selected by the same cost metric used for the selection of node $i$ in Step 1. The cost metric \cite{ref15,ref16} used in the opportunistic node selection procedure in Steps 1-2 ensures the selection of those nodes that can preserve the entangled quantum states with the highest fidelity in their local quantum memories. From a given distributing set ${\rm {\mathcal D}}$, only one node is selected in each iteration step. The cost function will be clarified later in the algorithm. 

Finally, Step 3 provides an iteration to reach from the source node to the target node.

\begin{center}
\begin{figure*}[!h]
%\vspace{-0.5cm}
\begin{center}
\includegraphics[angle = 0,width=0.9\linewidth]{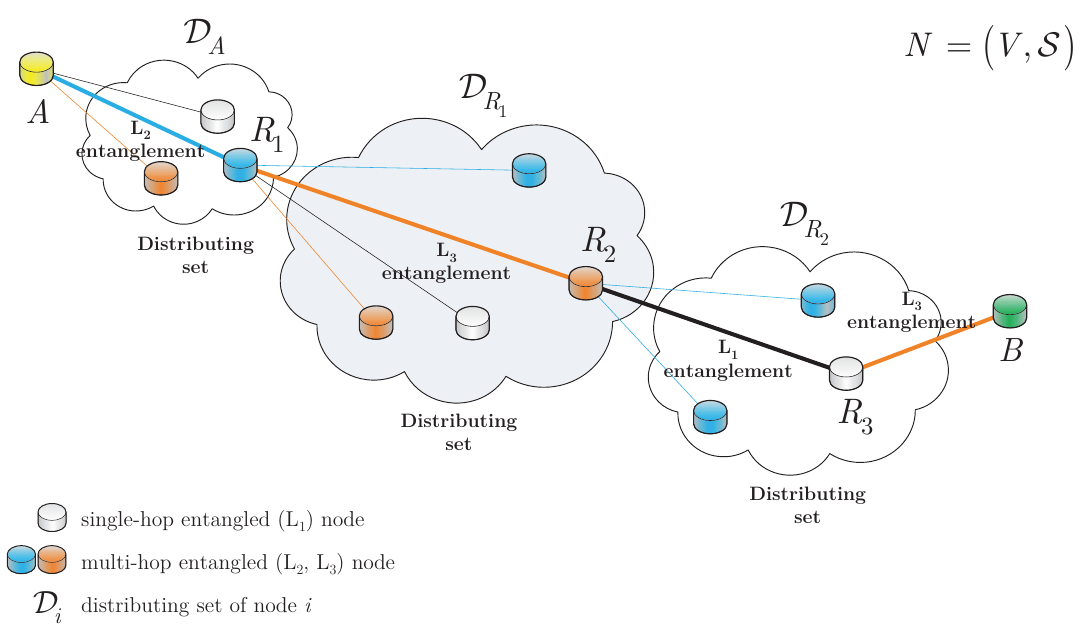}
\caption{Opportunistic entanglement distribution in a quantum Internet setting, $N=\left(V,{\rm {\mathcal S}}\right)$. The entangled contacts of transmitter node $A$ define the distributing set ${\rm {\mathcal D}}_{A} $. From set ${\rm {\mathcal D}}_{A} $, a quantum repeater node $R_{1} $ is selected. Node $R_{1} $ shares ${\rm L}_{2} $-level entanglement with $A$. The entangled contacts of the repeater node $R_{1} $ define the distributing set ${\rm {\mathcal D}}_{R_{1} } $. The iteration is repeated until target node $B$ is reached via a quantum repeater node. In this network setting, $B$ is reached via node $R_{3} $ from set ${\rm {\mathcal D}}_{R_{2} } $, where $R_{3} $ shares ${\rm L}_{3} $-level entanglement with $B$. Applying the opportunistic entanglement distribution for the nodes of the network, a path between $A$ and $B$ is selected (depicted in bold).} 
 \label{fig1}
 \end{center}
\end{figure*}
\end{center}

\fref{fig1} illustrates the method of opportunistic entanglement distribution in a quantum repeater network $N$. A given distributing set ${\rm {\mathcal D}}_{i} $ of a node $i$ can contain several different levels ${\rm L}_{l} $ of entangled contacts. From a distributing set ${\rm {\mathcal D}}_{i} $, only one repeater node is selected in each level of iteration. Those quantum nodes that share ${\rm L}_{1} $-level entanglement are referred to as single-hop entangled nodes, while the other entangled nodes are referred to as multi-hop entangled nodes.

\section{Results}
\label{sec4}
\subsection{Parameterization}

Let ${\rm {\mathcal D}}_{i} $ be a distributing set of a node $i$, and let $p_{i,j}^{F^{{\rm *}} } $ be the probability of the shared entanglement stored in the quantum memories of nodes $\left(i,j\right)$, where $j\in {\rm {\mathcal D}}_{i} $ identifies a repeater node $R_{j} $ from the distributing set ${\rm {\mathcal D}}_{i} $, with a fidelity $F^{{\rm *}} \ge F_{crit} $, where $F_{crit} $ is a critical lower bound on the fidelity of entanglement.

Then, let $p_{i,{\rm {\mathcal D}}_{i} }^{F^{{\rm *}} } $ be the probability that $F^{{\rm *}} \ge F_{crit} $ is satisfied between node $i$ and at least one repeater node from ${\rm {\mathcal D}}_{i} $. 

Using $p_{i,{\rm {\mathcal D}}_{i} }^{F^{{\rm *}} } $ and $p_{i,j}^{F^{{\rm *}} } $, a cost function $f\left(p_{i,{\rm {\mathcal D}}_{i} }^{F^{{\rm *}} } \right)$ can be defined \cite{ref15,ref16} as 
\begin{equation} \label{ZEqnNum989979} 
f\left(p_{i,{\rm {\mathcal D}}_{i} }^{F^{{\rm *}} } \right)=\frac{1}{p_{i,{\rm {\mathcal D}}_{i} }^{F^{{\rm *}} } } =\frac{1}{1-\prod _{j\in {\rm {\mathcal D}}_{i} }\left(1-p_{i,j}^{F^{{\rm *}} } \right) } .                                    
\end{equation} 
Similarly, with $p_{{\rm {\mathcal D}}_{i} ,B}^{F^{{\rm *}} } $ as the probability that there exists entanglement between a node $j\in {\rm {\mathcal D}}_{i} $ of ${\rm {\mathcal D}}_{i} $ and the target node $B$ with fidelity criterion $F^{{\rm *}} \ge F_{crit} $, a cost function between a repeater node $j\in {\rm {\mathcal D}}_{i} $ and a target node $B$ can be defined as
\begin{equation} \label{ZEqnNum801212} 
f\left(p_{{\rm {\mathcal D}}_{i} ,B}^{F^{{\rm *}} } \right)=\sum _{j\in {\rm {\mathcal D}}_{i} }\phi _{i,j} f\left(p_{j,B}^{F^{{\rm *}} } \right) ,                                          
\end{equation} 
where $p_{j,B}^{F^{{\rm *}} } $ is the probability of $F^{{\rm *}} \ge F_{crit} $ fidelity entangled contact between a given $j\in {\rm {\mathcal D}}_{i} $ and $B$, while $\phi _{i,j} $ is the probability that a given repeater node $j$ is selected from ${\rm {\mathcal D}}_{i} $ for $i$, defined \cite{ref16} as
\begin{equation} \label{ZEqnNum467938} 
\phi _{i,j} =\frac{p_{i,j}^{F^{{\rm *}} } \prod _{k=1}^{j-1}\left(1-p_{i,k}^{F^{{\rm *}} } \right) }{1-\prod _{j\in {\rm {\mathcal D}}_{i} }\left(1-p_{i,j}^{F^{{\rm *}} } \right) } ,                                              
\end{equation} 
where 
\begin{equation} \label{4)} 
\sum _{i}\phi _{i,j}  =1.                                                   
\end{equation} 
From equations \eqref{ZEqnNum989979} and \eqref{ZEqnNum801212}, the cost function between a quantum node $i$ and a target node $B$ at a distributing set ${\rm {\mathcal D}}_{i} $, such that $F^{{\rm *}} \ge F_{crit} $ holds for the fidelity of all entangled contacts from $i$ to $B$, is therefore 
\begin{equation} \label{5)} 
f\left(p_{i,B}^{F^{{\rm *}} } \right)=f\left(p_{i,{\rm {\mathcal D}}_{i} }^{F^{{\rm *}} } \right)+f\left(p_{{\rm {\mathcal D}}_{i} ,B}^{F^{{\rm *}} } \right).                                       
\end{equation} 
The $p_{i,j}^{F^{{\rm *}} } $ probabilities depend on the actual state of the quantum memory (particularly, on the $\varepsilon _{i} $ error probability of the quantum memory of node $i$) and on the $F$ fidelity of the stored entanglement. Parameters $\varepsilon $ and $F$ are time-varying in our model, which is denoted by $\varepsilon \left(t\right)$ and $F\left(t\right)$, where $t$ refers to the storage time in quantum memory. Time $t_{0} $ refers to the start of the storage of a quantum system in quantum memory. 

Let $\varepsilon _{i,j} \left(t_{0} +\Delta t\right)$ identify the error probability of quantum memories in nodes $\left(i,j\right)$ such that $\varepsilon _{i,j} \in \left[0,\varepsilon _{crit} \right]$ holds, which allows storing an $F^{{\rm *}} \ge F_{crit} $ fidelity entanglement in the nodes after $\Delta t$ time from start time $t_{0} $. Therefore, at a $\varepsilon _{crit} $ critical upper bound on the quantum memories, 
\begin{equation} \label{6)} 
\varepsilon _{i,j} \left(t_{0} +\Delta t\right)=\left|\zeta _{i} \left(t_{0} +\Delta t\right)-\zeta _{j} \left(t_{0} +\Delta t\right)\right|\le \varepsilon _{crit}  
\end{equation} 
holds, where $\zeta _{i} \left(t_{0} +\Delta t\right)$ characterizes the change of the error probability of the quantum memory of node $i$ at $t_{0} +\Delta t$, as in
\begin{equation} \label{7)} 
\zeta _{i} \left(t_{0} +\Delta t\right)=\zeta _{i} \left(t_{0} \right)+\varphi _{i} \left(t_{0} ,\Delta t\right),                                   
\end{equation} 
where $\zeta _{i} \left(t_{0} \right)$ is the $\varepsilon _{i} $ error probability of the quantum memory of node $i$ at $t_{0} $, while $\varphi _{i} \left(t_{0} ,\Delta t\right)$ is the change of $\varepsilon _{i} $ from $t_{0} $ to $t_{0} +\Delta t$.  

Let $\eta _{i} \left(t\right)=\left(\varepsilon _{i} \left(t\right),1-F_{i} \left(t\right)\right)$ identify the $\varepsilon _{i} \left(t\right)$ quantum memory error probability and the $F_{i} \left(t\right)$ stored entanglement fidelity in node $i$ at time $t$. Then, the $d_{i,j} $ distance between $\eta _{i} $ and $\eta _{j} $ in ${{\mathbb{R}}^{2}}$ identifies a $d_{i,j} \in \left[0,d_{\max } \right]$, where $d_{\max } $ is the maximal allowable distance in ${{\mathbb{R}}^{2}}$, which forms an upper bound to the distances for which $p_{i,j}^{F^{{\rm *}} } >0$ holds, as
\begin{equation} \label{8)} 
\begin{split}
   {{d}_{i,j}}\left( {{t}_{0}} \right)&=\left| {{\eta }_{i}}\left( {{t}_{0}} \right)-{{\eta }_{j}}\left( {{t}_{0}} \right) \right| \\ 
 & =\sqrt{{{\left( {{\varepsilon }_{i}}\left( {{t}_{0}} \right)-{{\varepsilon }_{j}}\left( {{t}_{0}} \right) \right)}^{2}}+{{\left( \left( 1-{{F}_{i}}\left( {{t}_{0}} \right) \right)-\left( 1-{{F}_{j}}\left( {{t}_{0}} \right) \right) \right)}^{2}}},  
\end{split}
\end{equation} 
and
\begin{equation} \label{ZEqnNum292706} 
d_{i,j} \left(t_{0} +\Delta t\right)=\left|\eta _{i} \left(t_{0} +\Delta t\right)-\eta _{j} \left(t_{0} +\Delta t\right)\right|\le d_{\max } ,                       
\end{equation} 
where 
\begin{equation} \label{10)} 
\eta _{i} \left(t_{0} +\Delta t\right)=\eta _{i} \left(t_{0} \right)+\Omega _{i} \left(t_{0} ,\Delta t\right),                                   
\end{equation} 
where 
\begin{equation} \label{11)} 
{{\eta }_{i}}\left( {{t}_{0}} \right)=\left( \begin{matrix}
   {{\zeta }_{i}}\left( {{t}_{0}} \right)  \\
   1-{{F}_{i}}\left( {{t}_{0}} \right)  \\
\end{matrix} \right)
\end{equation} 
and 
\begin{equation} \label{12)} 
{{\Omega }_{i}}\left( {{t}_{0}},\Delta t \right)=\left( \begin{matrix}
   {{\varphi }_{i}}\left( {{t}_{0}},\Delta t \right)  \\
   1-{{F}_{i}}\left( {{t}_{0}},\Delta t \right)  \\
\end{matrix} \right).
\end{equation} 
The fidelities $F_{i} \left(t_{0} \right)$ and $F_{i} \left(t_{0} ,\Delta t\right)$ refer to the fidelities of stored entanglement in a node $i$ at $t_{0} $ and $t_{0} ,\Delta t$, respectively, with relation for nodes $\left(i,j\right)$ 
\begin{equation} \label{13)} 
1-F_{i,j} \left(t_{0} +\Delta t\right)=\left|\left(1-F_{i} \left(t_{0} +\Delta t\right)\right)-\left(1-F_{j} \left(t_{0} +\Delta t\right)\right)\right|\le 1-F_{\Delta } ,         
\end{equation} 
where $F_{i,j} $ is the difference of entanglement fidelities $F_{i} $ and $F_{j} $, while $F_{\Delta } $ is an upper bound on the fidelity difference.

From equation \eqref{ZEqnNum292706}, the following relation holds for $p_{i,j}^{F^{{\rm *}} } $:
\begin{equation} \label{14)} 
p_{i,j}^{F^{{\rm *}} } =\left\{\begin{array}{l} {p_{i,j}^{F^{{\rm *}} } >0,{\rm \; }\text{if}{\rm \; }d_{i,j} \left(t_{0} +\Delta t\right)\le d_{\max }, } \\ {p_{i,j}^{F^{{\rm *}} } =0,{\rm otherwise.\; \; \; \; \; \; \; \; \; \; \; \; \; \; \; \; \; \; \; }} \end{array}\right.  
\end{equation} 

\begin{center}
\begin{figure*}[!h]
%\vspace{-0.5cm}
\begin{center}
\includegraphics[angle = 0,width=0.9\linewidth]{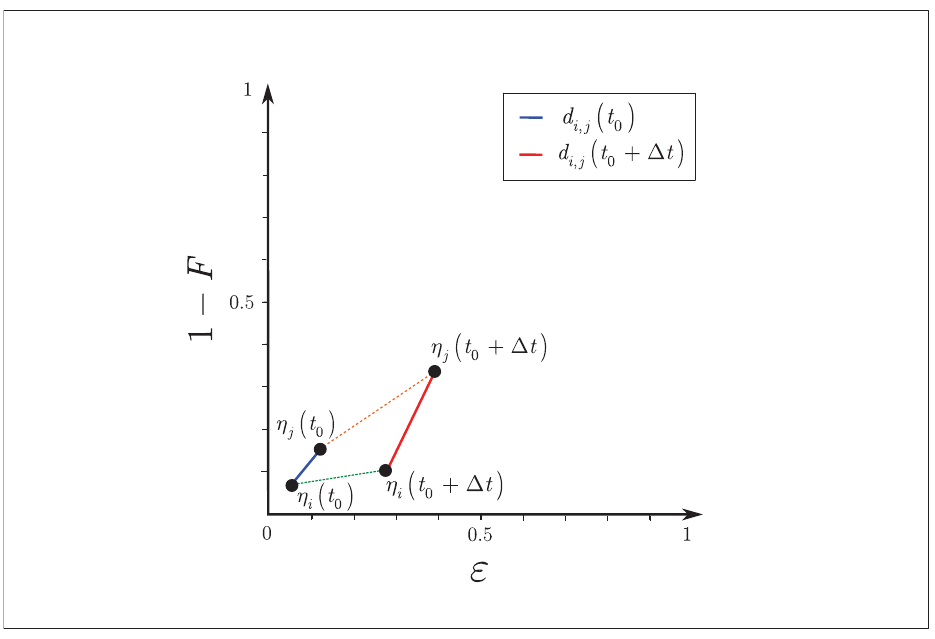}
\caption{Evolution of the $d_{i,j} \left(t_{0} \right)$ and $d_{i,j} \left(t_{0} +\Delta t\right)$ distances in ${{\mathbb{R}}^{2}}$ with respect to a given node pair $\left(i,j\right)$; $\varepsilon $ is the error probability of the local quantum memory and $F$ is the fidelity of the stored entanglement in the local quantum memory. The initial states (storage starting at $t_{0} $) of the nodes $\left(i,j\right)$ are identified by $\eta _{i} \left(t_{0} \right)$ and $\eta _{j} \left(t_{0} \right)$, the current states at time $t_{0} +\Delta t$ of the nodes are $\eta _{i} \left(t_{0} +\Delta t\right)$ and $\eta _{j} \left(t_{0} +\Delta t\right)$. As $d_{i,j} \left(t_{0} +\Delta t\right)>d_{\max } $, where $d_{\max } $ is a threshold, the yielding probability is $p_{i,j}^{F^{{\rm *}} } =0$.} 
 \label{fig2}
 \end{center}
\end{figure*}
\end{center}

\fref{fig2} illustrates the evolution of the error probabilities of the local quantum memories and the fidelities of the stored entangled states. Time $t_{0} $ refers to the start of storage in the quantum memory and time $t_{0} +\Delta t$ is the current time. The distance $d_{i,j} \left(t_{0} +\Delta t\right)$ measures the difference of $\eta _{i} \left(t_{0} +\Delta t\right)$ and $\eta _{j} \left(t_{0} +\Delta t\right)$ of a node pair $\left(i,j\right)$. From the time evolution of the error probability of the quantum memory, it follows that $\varepsilon \left(t_{0} +\Delta t\right)>\varepsilon \left(t_{0} \right)$, and $F\left(t_{0} +\Delta t\right)<F\left(t_{0} \right)$ holds for the time evolution of the fidelity of the stored entanglement.

The normalized increase of distance $d_{i,j} $ between nodes therefore yields a quantity $\omega _{i,j} $, $\omega _{i,j} \in \left[0,1\right]$, which characterizes the usability of a stored entanglement in $\left(i,j\right)$ from $t_{0} $ to $t_{0} +\Delta t$, as 
\begin{equation} \label{ZEqnNum538722} 
{{\omega }_{i,j}}=1-\frac{1}{{{d}_{\max }}}\min \left( \sqrt{\begin{split}
  & {{\left( {{\varphi }_{i}}\left( {{t}_{0}},{{t}_{0}}+\Delta t \right)-{{\varphi }_{j}}\left( {{t}_{0}},{{t}_{0}}+\Delta t \right) \right)}^{2}} \\ 
 & +{{\left( 1-{{F}_{i}}\left( {{t}_{0}},{{t}_{0}}+\Delta t \right)-\left( 1-{{F}_{j}}\left( {{t}_{0}},{{t}_{0}}+\Delta t \right) \right) \right)}^{2}} \\ 
\end{split}},{{d}_{\max }} \right).
\end{equation} 
As $\omega _{i,j} =1$, if no change occurs in the initial distance $d_{i,j} \left(t_{0} \right)$ in ${{\mathbb{R}}^{2}}$, thus
\begin{equation} \label{16)} 
d_{i,j} \left(t_{0} +\Delta t\right)\approx d_{i,j} \left(t_{0} \right),                                            
\end{equation} 
while $\omega _{i,j} =0$, if 
\begin{equation} \label{17)} 
d_{i,j} \left(t_{0} +\Delta t\right)>d_{\max } ,                                             
\end{equation} 
so the yielding relation between $p_{i,j}^{F^{{\rm *}} } $ and $\omega _{i,j} $ is 
\begin{equation} \label{18)} 
p_{i,j}^{F^{{\rm *}} } =\left\{\begin{array}{l} {p_{i,j}^{F^{{\rm *}} } =\max ,{\rm \; }\text{if}{\rm \; }\omega _{i,j} =1,{\rm \; \; \; \; \; \; \; \; \; \; }} \\ {0<p_{i,j}^{F^{{\rm *}} } <\max ,{\rm \; }\text{if}{\rm \; }\omega _{i,j} <1,{\rm \; \; \; \; }} \\ {p_{i,j}^{F^{{\rm *}} } =0,{\rm \; }\text{if}{\rm \; }\omega _{i,j} =0.{\rm \; \; \; \; \; \; \; \; \; \; \; \; \; \; }} \end{array}\right.  
\end{equation} 
Equation \eqref{ZEqnNum538722} also characterizes the future behavior of the quantum memories in the nodes, i.e., the predictability of the error model of the quantum memories after a given time after beginning storage. The highest values of $\omega _{i,j} $ are therefore assigned to those memory units for which the error probabilities and the entanglement fidelities evolve by a given pattern.     

For a given path ${\rm {\mathcal P}}$ with $A_{\delta ,U_{k} } $ and $B_{\delta ,U_{k} } $ as the source and target nodes, respectively, associated with a demand $\delta $ of user $U_{k} $, $k=1,\ldots ,K$, where $K$ is the number of users; thus, the overall usability coefficient $\omega _{A_{\delta ,U_{k} } ,B_{\delta ,U_{k} } } $ for ${\rm {\mathcal P}}$ is 
\begin{equation} \label{19)} 
\omega _{A_{\delta ,U_{k} } ,B_{\delta ,U_{k} } } =\prod _{\left(i,j\right)\in {\rm {\mathcal P}}}\omega _{i,j}  .                                             
\end{equation} 
To determine those quantum nodes for which both $\omega _{i,j} $ \eqref{ZEqnNum538722} and $p_{i,j}^{F^{{\rm *}} } $ are high, a redefined cost function, $c_{i,j} $, can be defined for a given $\left(i,j\right)$ as 
\begin{equation} \label{ZEqnNum151929} 
c_{i,j} =\frac{1}{p_{i,j}^{F^{{\rm *}} } \omega _{i,j} } , 
\end{equation} 
which assigns the lowest cost to those node pairs for which $\omega _{i,j} $ and $p_{i,j}^{F^{{\rm *}} } $ are high. 

The remaining quantities from equations \eqref{ZEqnNum989979} and \eqref{ZEqnNum467938} can therefore be rewritten as
\begin{equation} \label{21)} 
c_{i,{\rm {\mathcal D}}_{i} } =\frac{1}{1-\prod _{j\in {\rm {\mathcal D}}_{i} }\left(1-p_{i,j}^{F^{{\rm *}} } \omega _{i,j} \right) } ,                                          
\end{equation} 
and
\begin{equation} \label{22)} 
\phi '_{i,j} =\frac{p_{i,j}^{F^{{\rm *}} } \omega _{i,j} \prod _{k=1}^{j-1}\left(1-p_{i,k}^{F^{{\rm *}} } \omega _{i,k} \right) }{1-\prod _{j\in {\rm {\mathcal D}}_{i} }\left(1-p_{i,j}^{F^{{\rm *}} } \omega _{i,j} \right) } ,                                          
\end{equation} 
which yields the redefined cost of equation \eqref{ZEqnNum801212} as
\begin{equation} \label{23)} 
c_{{\rm {\mathcal D}}_{i} ,B} =\sum _{j\in {\rm {\mathcal D}}_{i} }\phi '_{i,j} f\left(p_{j,B}^{F^{{\rm *}} } \right) .                                            
\end{equation} 
The total cost between $i$ and $B$ such that $F^{{\rm *}} \ge F_{crit} $ holds for all entangled contacts between quantum nodes $i$ and $B$ is therefore \cite{ref15,ref16}
\begin{equation} \label{24)} 
c_{i,B} =c_{i,{\rm {\mathcal D}}_{i} } +c_{{\rm {\mathcal D}}_{i} ,B} .                                              
\end{equation} 

\subsection{Opportunistic Entanglement Distribution}

The aim of the opportunistic entanglement distribution algorithm ${\rm {\mathcal A}}_{O} $ is to determine a shortest path ${\rm {\mathcal P}}^{{\rm *}} $ with respect to our cost function. The shortest path ${\rm {\mathcal P}}^{{\rm *}} $ contains those node pairs for which $\omega _{i,j} $ and $p_{i,j}^{F^{{\rm *}} } $ are maximal, and therefore the resulting cost function \eqref{ZEqnNum151929} is the lowest in a quantum network $N$. Thus, the algorithm finds the repeater nodes for entanglement distribution by maximizing the usability of stored entanglement.

Some preliminary notations for the algorithm are as follows. Let $A_{\delta ,U_{k} } $ and $B_{\delta ,U_{k} } $ be the source and target nodes, respectively, associated with a demand $\delta $ of user $U_{k} $. The algorithm selects those nodes that provide the lowest cost with respect to $c_{i,j} $ for a given $\left(i,j\right)$ to distribute entanglement from $A_{\delta ,U_{k} } $ to $B_{\delta ,U_{k} } $. Assume that for a set $S'$ of nodes there exists a path to $B_{\delta ,U_{k} } $ in a quantum network $N=\left(V,S\right)$. Let $\tilde{S}$ refer to a set of nodes for which the shortest path to $B_{\delta ,U_{k} } $ is not yet determined. 

The ${\rm {\mathcal A}}_{O} $ algorithm of the minimal-cost opportunistic entanglement distribution is detailed in Algorithm 1.

\setcounter{algocf}{0}
\begin{algo}
  \DontPrintSemicolon
\caption{${\rm {\mathcal A}}_{O}$: Minimal-cost opportunistic entanglement distribution}

\textbf{Step 1}. For all quantum node $i\in V$, initial cost $f(p_{i,B_{\delta ,U_{k} } }^{F^{{\rm *}} })=\infty $ and ${\rm {\mathcal D}}_{i} =0$, where $f(p_{i,B_{\delta ,U_{k} } }^{F^{{\rm *}} })$ is the cost of a path from $i$ to $B_{\delta ,U_{k} } $, while ${\rm {\mathcal D}}_{i} =0$ is the distributing set associated to node $i$ to reach $B_{\delta ,U_{k} } $. Set $\tilde{S}=V$ with cost $f(p_{i\in \tilde{S},B_{\delta ,U_{k} } }^{F^{{\rm *}} })=0$, and set $S'=\emptyset $. 

\textbf{Step 2}. Determine the final cost of the path with respect to a node $\Phi _{n} $ from  set $\tilde{S}$ with minimal cost $\tilde{f}(p_{\Phi _{n} ,B_{\delta ,U_{k} } }^{F^{{\rm *}} })$, where $\tilde{f}\left( \cdot  \right)$ is an upper bound on the cost of the shortest path from $\Phi _{n} $ to $B_{\delta ,U_{k} } $, while node $\Phi _{n} $ is determined as $\Phi _{n} =\mathop{\min }\limits_{z\in \tilde{S}} f(p_{z,B_{\delta ,U_{k} } }^{F^{{\rm *}} })$, where $z$ is a node from set $S'$. 

\textbf{Step 3}. Set $S'=S'\bigcup \left\{\Phi _{n} \right\}$, and for each $\left(i,\Phi _{n} \right)$ set ${\rm {\mathcal D}}={\rm {\mathcal D}}_{i} \bigcup \left\{\Phi _{n} \right\}$, where ${\rm {\mathcal D}}$ is a distributing set.

\textbf{Step 4}. If $f(p_{i,B_{\delta ,U_{k} } }^{F^{{\rm *}} })>\tilde{f}(p_{\Phi _{n} ,B_{\delta ,U_{k} } }^{F^{{\rm *}} })$, update the cost of node $i$ as $f(p_{i,B_{\delta ,U_{k} } }^{F^{{\rm *}} })=c_{i,{\rm {\mathcal D}}_{i} } +c_{{\rm {\mathcal D}}_{i} ,B_{\delta ,U_{k} } } $, and set ${\rm {\mathcal D}}_{i} ={\rm {\mathcal D}}$. 

\textbf{Step 5}. Repeat steps 2-4, until $\tilde{S}\ne \emptyset $. 

\textbf{Step 6}. Output the minimal cost path ${\rm {\mathcal P}}^{{\rm *}} $ between $i_{\delta ,U_{k} } $ and $B_{\delta ,U_{k} } $.

\end{algo}

\subsection{Computational Complexity}

The computational complexity of the minimal-cost ${\rm {\mathcal A}}_{O} $ opportunistic entanglement distribution is as follows. 

Let $N=\left(V,E\right)$ be a quantum repeater network with $\left|V\right|$ quantum nodes. Applying a ${\rm {\mathcal L}}$ logarithmic search \cite{ref17} to find a node with an actual minimal cost requires at most 
\begin{equation} \label{ZEqnNum468697} 
{\rm {\mathcal O}}\left(\log \left|V\right|\right) 
\end{equation} 
steps, via ${\rm {\mathcal O}}\left(\log \left|V\right|\right)$ comparisons. 

Since the number of nodes is $\left|V\right|$, setting the final path cost for all quantum nodes requires at most 
\begin{equation} \label{ZEqnNum530807} 
{\rm {\mathcal O}}\left(\left|V\right|\right) 
\end{equation} 
steps. 

From \eqref{ZEqnNum468697} and \eqref{ZEqnNum530807} follows straightforwardly that the complexity of the minimal-cost opportunistic entanglement distribution algorithm is bounded above by
\begin{equation} \label{27)} 
{\rm {\mathcal O}}\left(\left|V\right|\log \left|V\right|\right).                                                   
\end{equation} 

\section{Conclusions}
\label{sec5}
Here we defined a method for entanglement distribution in the quantum Internet. Our method utilizes distributing sets for quantum nodes, which can preserve quantum entanglement with the highest fidelity in their local quantum memories. The algorithm is opportunistic, since in each iteration step a node is selected from a distributing set that can provide optimal conditions. The cost function includes the utilization of the evolution of the error model of the local quantum memories, and the fidelities of the stored entangled states. A usability parameter quantifies the predictability of the evolution of the error model, and of the evolution of the entanglement fidelity. The computational complexity of the method is moderate, which allows for direct application in experimental quantum Internet scenarios and in long-distance quantum communications.

%\section*{Statements}
%\subsection*{Ethics statement}
%This work did not involve any active collection of human data.
%\subsection*{Data accessibility statement}
%This work does not have any experimental data.
%\subsection*{Competing financial interests statement}
%We have no competing financial interests.
%\subsection*{Competing interests statement}
%We have no competing interests.
%\subsection*{Funding}
%No relevant funding. 
%\subsection*{Authors’ contributions}
%L.GY. designed the protocol and wrote the manuscript. L.GY. and S.I. analyzed the results. All authors reviewed the manuscript.

\section*{Acknowledgements}
This work was partially supported by the European Research Council through the Advanced Fellow Grant, in part by the Royal Society’s Wolfson Research Merit Award, in part by the Engineering and Physical Sciences Research Council under Grant EP/L018659/1, by the Hungarian Scientific Research Fund - OTKA K-112125, and by the National Research Development and Innovation Office of Hungary (Project No. 2017-1.2.1-NKP-2017-00001), and in part by the BME Artificial Intelligence FIKP grant of EMMI (BME FIKP-MI/SC).

\end{document}